Numerical Simulation Informed Rapid Cure Process Optimization of Composite Structures using Constrained Bayesian Optimization

**Authors:** Madhura Limaye[1, 2#], Yezhuo Li[3#], Qiong Zhang[3], Gang Li[2*]

[1]Manufacturing Science Division, Oak Ridge National Laboratory, Knoxville, TN, 37932, USA
[2]Department of Mechanical Engineering, Clemson University, Clemson, SC - 29631, USA
[3]School of Mathematical and Statistical Sciences, Clemson University, Clemson, SC - 29634, USA

*Author to whom correspondence should be addressed. Electronic mail: gli@clemson.edu
# Authors contributed equally to this work

**Abstract**

The present study aimed to solve the cure optimization problem of laminated composites through a statistical approach. The approach consisted of using constrained Bayesian Optimization (cBO) along with a Gaussian process model as a surrogate to rapidly solve the cure optimization problem. The approach was implemented to two case studies including the cure of a simpler flat rectangular laminate and a more complex L-shaped laminate. The cure optimization problem with the objective to minimize cure induced distortion was defined for both case studies. The former case study was two-variable that is used two cure cycle parameters as design variables and was constrained to achieve full cure, while the latter was four-variable and had to satisfy constraints of full cure as well as other cure cycle parameters. The performance of cBO for both case studies was compared to the traditional optimization approach based on Genetic Algorithm (GA). The comparison of results from GA and cBO including deformation and final degree of cure showed significant agreement (error < 4%). The computational efficiency of cBO was calculated by comparing the convergence steps for GA (>1000) and cBO (<50). The computational efficiency of cBO for all optimization cases was found to be > 96%. The case studies conclude that cBO is promising in terms of computational time and accuracy for solving the cure optimization problem.

***Keywords:*** *Cure behavior, Bayesian Optimization, Process-induced deformation (PID), Thermoset composites, Finite element analysis*

## 1. Introduction

Thermoset based fiber reinforced composite laminates are popularly processed by the cure process to produce structural parts. These structural parts have found extensive applications in aerospace and automotive industries (McIlhagger, Archer, and McIlhagger 2020; Pradeep et al. 2024). A typical input to the cure process is a temperature and pressure cycle also commonly known as a cure cycle. The cure cycle initiates and enables completion of the cure reactions of the thermoset prepreg. The cure cycle also enables laminate consolidation and void reduction by excessive resin bleed-out and impregnation of fibers. However, the cure processing of composite structures often suffers from residual stress inducement through internal and external sources. The main sources of residual stress development in a composite laminate include a) mismatch of thermal expansion coefficients (CTE) at micro-level (fiber-resin interaction) and macro-level (ply-to-ply interaction), b) cure shrinkage of resin, c) temperature gradient through laminate thickness and, d) interaction between the composite laminate and the tool. In particular, the two sources of residual stress development that have received considerable attention from researchers and industry alike are mismatch of CTE (thermal effects) and the cure shrinkage of resin (cure shrinkage effects) (Gopal, Adali, and Verijenko 2000; Kravchenko, Kravchenko, and Pipes 2017; Sreekantamurthy et al. 2016; Genidy, Madhukar, and Russell 2000; Madhukar, Genidy, and Russell 2000; Russell et al. 2000; S.R. White and Hahn 1993).

The residual stresses induced through these sources severely compromises the strength and adversely affect the performance of the composite laminate (Hahn 1976; Agius et al. 2016; Zhao, Warrior, and Long 2006; C. Li et al. 2014; S.R. White and Hahn 1993). For example, these process-induced residual stresses have shown to cause matrix cracking which compromised its strength before mechanical loading (Hahn 1976; S.R. White and Hahn 1993). In addition, cure process induced residual stresses were shown to have significant effect on the tensile strength of the matrix material (C. Li et al. 2014). Further, these residual stresses cause deformations in the laminate referred to as the process induced deformation (PID), and lead to deviations from the nominal dimensions. Such distorted laminate parts in assembly then give rise to mounting stresses (Ersoy et al. 2010a; Kravchenko, Kravchenko, and Pipes 2016; Gigliotti, Wisnom, and Potter 2003; Russell et al. 2000). Thus, minimizing or eliminating these process-induced residual stresses/deformation plays a key role in the manufacturing of high-quality composite structures. Experimental and numerical studies have shown that these process-induced stresses are directly influenced by the cure cycle the thermoset prepregs are subjected to (Kravchenko, Kravchenko, and Pipes 2017; Sarrazin et al. 1995; Madhukar, Genidy, and Russell 2000; S.R. White and Hahn 1993). As a result, determination of an optimal cure cycle is critical to reduce residual stresses and restrict deformation within prescribed tolerances. In the past, the determination of an optimal cure cycle was largely based on trial-and-error experimental methods (Purslow and Childs 1986; Madhukar, Genidy, and Russell 2000; Genidy, Madhukar, and Russell 2000; Russell et al. 2000; S.R. White and Hahn 1993). Such experimentally driven techniques are both expensive and time-consuming especially for large complex structures. Hence, there is a pressing need for adopting computational methods that adequately capture relevant physics of the cure problem to deliver optimized solutions for desired part quality.

Computational cure process models implemented with the Finite Element Method (FEM) were extensively employed to predict process induced stresses/deformation (Svanberg and Holmberg 2004; Bapanapalli and Smith 2005; Zeng and Raghavan 2010; Ersoy et al. 2010b; Ding et al. 2016; Sun et al. 2017; Takagaki, Minakuchi, and Takeda 2017; Ding et al. 2017; Bellini and Sorrentino 2018; Benavente et al. 2018; D. Li et al. 2018; Chen and Zhang 2019; Liu et al. 2021). As the thermo-chemo-mechanical behavior of the composite laminate is largely controlled by the curing resin, different constitutive models for capturing the resin behavior during cure were developed. Among those, the purely elastic model could only predict stresses in the cool-down phase while the most realistic viscoelastic model could accurately determine the history-dependent residual stresses (Scott R. White and Kim 1998). However, such complex model requires extensive material characterization along with expensive computation. Thus, simplified models such as CHILE, and path-dependent models were proposed to determine process induced stresses/deformation in more computationally efficient manner. Alternatively, surrogate models were generated by sampling from the complex cure process model predictions to save computational time and cost. These surrogate models were also used to conduct sensitivity analysis and determine most influential parameters. As a next step, the cure process models were coupled with optimization procedures to determine optimum process parameters that reduce the stresses/deformation. For example, Shah et al. constructed response surface from cure process model and used it to run genetic algorithm cases to minimize PID while achieving full cure (Shah et al. 2018). Szarski et al. trained Reinforcement learning (RL) models to adaptively control cure profile and used learnings of the RL model to conduct tooling optimization via Bayesian Optimization (Szarski and Chauhan 2021). In another study, the traditional Genetic Algorithm (GA) was improved by Li X et al. such that the 2-step GA improved the search strategy by first identifying the probable optimum spaces and next zooming into each space, thus achieving computational efficiency and better accuracy (X. Li et al. 2021). In another independent study, the authors (Hui et al. 2022) used a multi-scale prediction strategy to simultaneously minimize laminate temperature gradients, residual stresses and cure time. The cure model with temperature dynamics evaluated at macro-scale and residual stresses calculated at RVE micro-scale was integrated with GA optimization NSGA-II to obtain optimum cure cycle parameters. In yet another study, the authors (Wang et al. 2022) conducted sensitivity analysis and demonstrated relationships between cure parameters and the target variables. An RBF surrogate model trained from samples of viscoelastic cure model was first used for the SA and next for running optimization cases using a combination of the influential parameters. They concluded that considering all cure cycle parameters and stress relaxation effects is necessary to achieve globally optimum cure cycle.

While many studies have focused on improving the numerical prediction accuracy through modified algorithms, complex equations and sophisticated surrogate models, others have attempted to solve a generalized multi-objective cure optimization problem considering several physical and structural parameters and presented a potential global solution. However, these studies do not consider underlying factors/mechanisms relating the cure cycle parameters to the resulting residual stress/deformation. Most studies have quantified residual stress/deformation combining thermal and cure shrinkage effects while considering isothermal cure cycles [92,98,104,106,112]. It has been pointed out in literature that the cure shrinkage effects, and thermal effects form a counteractive mechanism that could be effectively controlled through a modified multi-ramp cure cycle to reduce the PID [93,94]. Such a modified cure cycle consists of a non-isothermal ramp up to cure temperature and is hence referred to as a non-isothermal cure cycle in this study. In the present work, the cure optimization problem was solved by using such non-isothermal cure cycles to take full advantage of these counteracting effects. Further, the existing optimization methods need to be sped up to meet the development requirements of structures of larger size and higher complexities. For that purpose, a tradeoff between numerical accuracy and computational efficiency needs to be prioritized and a strategy needs to be developed to solve physically and geometrically diverse cure problems. This was addressed in the present work by replacing the elaborate global search methods with efficient search algorithm.

This work presents a constrained Bayesian Optimization (cBO) (Gardner et al. 2014) approach for solving the cure optimization problem and demonstrates its computational efficiency over other traditional derivative-free optimization algorithms. This was done by: (1) defining a cure optimization problem to minimize the internal residual stress development/ deformation during the cure process within the feasible range of the cure cycle parameters, (2) solving this cure optimization problem with two approaches: i) the traditional Genetic Algorithm (GA) NSGA-II, and ii) the constrained Bayesian Optimization (cBO) algorithm, (3) comparing the solution obtained from the two approaches on the metrics of numerical accuracy and computational efficiency. This comparative study was performed for two different cure scenarios in two separate case studies, first for the cure of a flat rectangular laminate and second for a L-shaped laminate. With a comparison of GA and BO for the two case studies, it was demonstrated that BO algorithm significantly improved computational efficiency ( > 96%) while maintaining the numerical accuracy (error < 4%). The rest of the manuscript is organized as follows. In Section 2, we describe the problem of cure process optimization and discuss a traditional genetic algorithm based black-box optimization approach to solve this problem. In Section 3, we describe the constraint Bayesian optimization approach. In Section 4, implementation of both optimization approaches to two case studies is discussed. Next, the results of the two optimization approaches implemented to two case studies are provided, compared and discussed in Section 5. Finally, conclusions are summarized in Section 6.

## 2. The curing process optimization problem

Fig 1 shows the workflow of the cure optimization problem. The initial design inputs were defined using Latin Hypercube sampling technique. The design evaluations were performed using a physics-based cure process model. The results of the evaluations that is the design outputs were provided to a black-box optimization toolbox that generated the design inputs for the next generation. The new generation design inputs were again evaluated with the cure process model. Accordingly, optimization loop was allowed to run for a defined number of generations. The cure optimization problem was defined as

$$\text{Objective: } Minimize(u)$$
$$\text{Contraints: } 1. DoC \geq 0.960$$
$$\text{Inputs: } t, T$$

The design inputs for the optimization problem were generated by controlling one or more points on the cure cycle in the time (t)-temperature (T) plane, as shown at the center of Fig 1. The objective of the optimization problem was to minimize $u$ which is the cure induced deformation. The constraint was defined such that the degree of cure ($DoC$) of the composite laminate should be greater than or equal to 0.96. The cure optimization problem was set up in ModeFRONTIER software. The cure process model was developed using a commercial FEA software ABAQUS along with a cure simulation tool COMPRO. The computational cure analysis was performed through a two-step simulation procedure. The two steps, namely the Thermo-chemical step and stress-deformation step, were sequentially coupled and the results of the first step are used as input to the subsequent step. The governing equations and material models of the cure process model setup are described in this section.

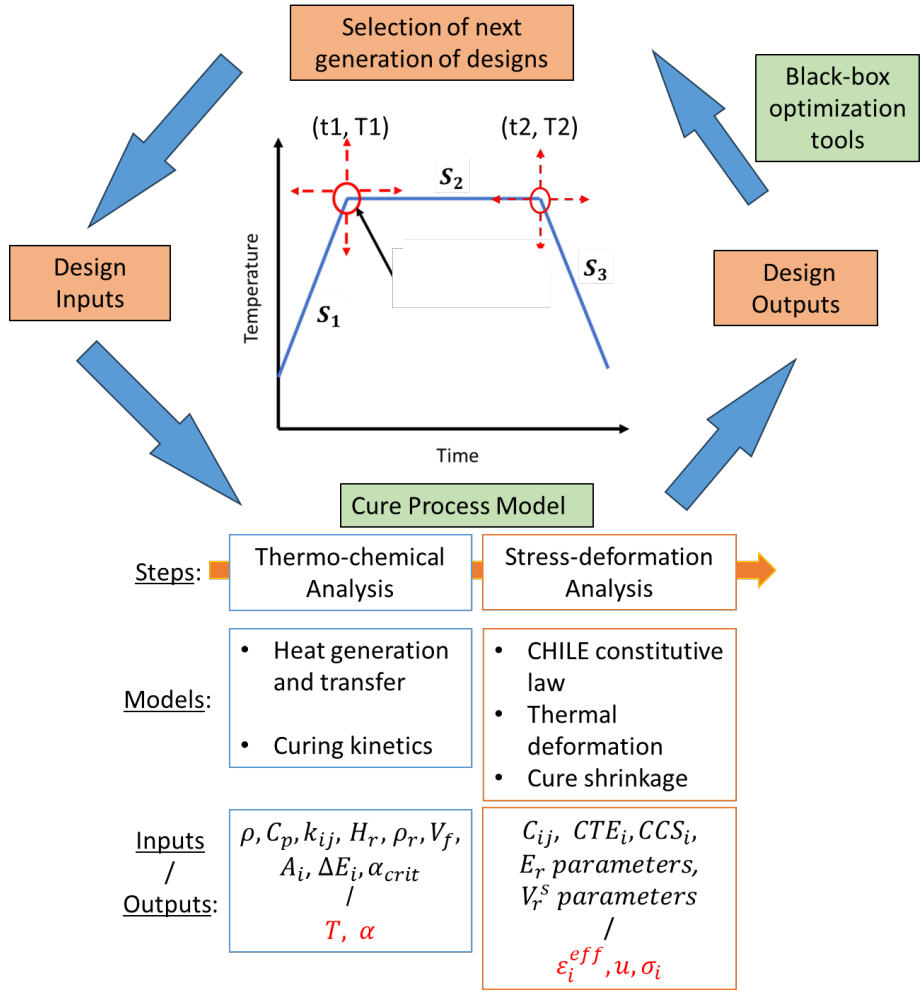

Figure 1. Workflow of the cure optimization problem with black-box optimization tools.

## 2.1 Heat generation and transfer model

The thermal model of cure process governed by Fourier's heat conduction equation with heat generation is given as

$$\frac{\partial}{\partial t}(\rho C_p T) = \nabla(\boldsymbol{k}\nabla T) + \dot{Q} \tag{1}$$

where $\rho$ is the density of the composite laminate, $C_p$ is the specific heat, $\boldsymbol{k}$ is the anisotropic thermal conductivity, which are determined at every time increment with micromechanical equations following rule of mixtures (Ma et al. 2015). $\dot{Q}$ is the resin heat generation rate which is given by

$$\dot{Q} = \frac{d\alpha}{dt}(1 - V_f)\rho_r H_R \tag{2}$$

where $\alpha$ is the degree of cure which is a measure of the extent of cross-linking in the thermosetting polymer, $\frac{d\alpha}{dt}$ is the cure rate calculated using a cure kinetic model, $V_f$ is the fiber volume fraction, $\rho_r$ is the resin density and $H_R$ is the resin heat of reaction/total heat evolved during the cure process.

## 2.2 Cure Kinetic model

The chemical model of cure process is defined by cure kinetic equations. Various kinetic models have been developed to describe the cure rate of resin systems ($\frac{d\alpha}{dt}$) as a function of the DoC and temperature, the selection of model depends on the cure behavior of a particular resin. Two types of kinetic models have been developed using the model fitting approach: phenomenological models (Boey et al. 2002; Morgan et al. 1997; Abbate et al. 1997; Goodwin 1993) and mechanistic models (Phelan and Sung 1997; Hopewell, George, and Hill 2000). The phenomenological models are based on empirical rate laws and do not incorporate the details of the resin reaction. The mechanistic models take into account the details of the reaction such as species, concentration and other factors. For the epoxy material used in the present study, a phenomenological model is fit to express the cure kinetics as follows (Woo, Loos, and Springer 1982)

$$\frac{d\alpha}{dt} = \begin{cases} (B_1 + \alpha B_2)(1-\alpha)(\alpha_{crit} - \alpha), & \alpha \leq 0.3 \\ B_3(1-\alpha), & 1 \geq \alpha \geq 0.3 \end{cases} \quad (3)$$

where $\alpha$ is the degree of cure, $B_i = A_i e^{\frac{\Delta E_i}{RT}}$ where $A_i, \Delta E_i$ are the pre-exponential factors and activation energies, $R$ is the universal gas constant and $T$ is the temperature. The values of the model parameters are provided in Appendix A2.

## 2.3 Mechanical constitutive model

The mechanical model for the cure process is governed by the resin constitutive model. The resin Youngs's modulus increases several orders of magnitude due to crosslinking reaction. This modulus development phenomenon is defined using a modified CHILE model in the present study. The CHILE model is such that ply properties are linear elastic for every time step and has been defined previously as a function of degree of cure $\alpha$ (Johnston 1997) and glass transition temperature $Tg$ (Liu et al. 2021). The modified CHILE model implemented in the present study is CHILE($\alpha$) such that the modulus development is assumed to occur from gelation upto vitrification of the resin. This is demonstrated with the help of a **P**olymerization-**G**elation-**V**itrification (PGV) plot shown in Fig 2. At any point on the laminate subjected to a cure cycle, The Gelation (G) is assumed to occur when viscosity reaches 100 Pa.s. for the 3501-6 resin, while Vitrification (V) is assumed to occur when the instantaneous $Tg$ exceeds the laminate temperature.

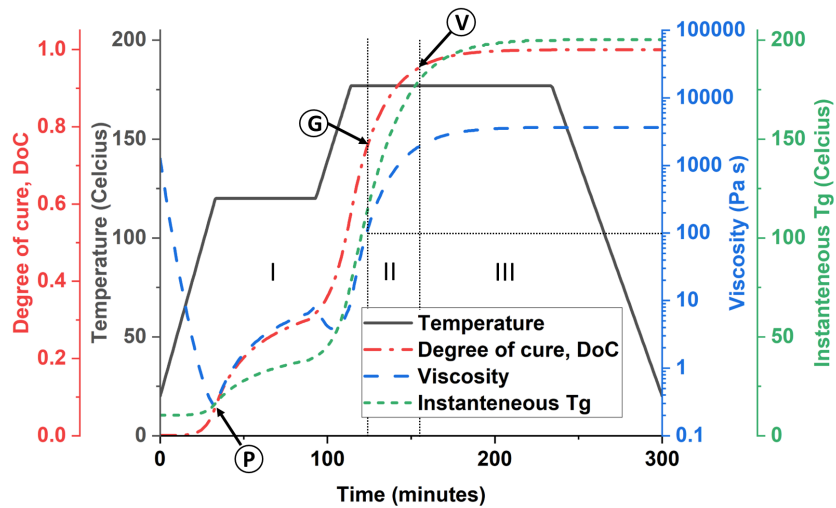

Figure 2. Property development for a laminate subjected to a two-step cure cycle (black curve) demonstrated through degree of cure profile (dash-dotted red curve), viscosity (dashed blue curve), and instantaneous glass transition temperature ($Tg$) profile (short dashed green curve). The cure cycle is divided into three phases: viscous phase (I), visco-elastic phase (II) and elastic phase (III).

The modified CHILE($\alpha$) is defined as:

$$E_r = \begin{cases} E_r^0, & \alpha \leq \alpha_1 \\ (1-\alpha_{mod})E_r^0 + \alpha_{mod}E_r^\infty + \gamma\alpha_{mod}(1-\alpha_{mod})(E_r^\infty - E_r^0), & \alpha_1 \leq \alpha \leq \alpha_2 \\ E_r^\infty, & \alpha > \alpha_2 \end{cases} \quad (4)$$

where $E_r^0$ is the modulus of the resin in the liquid phase and $E_r^\infty$ is the modulus in the solid or fully cured state. $\alpha_1$ and $\alpha_2$ are the DoC parameters for resin modulus development model. $\alpha_1$ is the degree of cure (DoC) corresponding to Point G in Fig 2, and $\alpha_2$ is the degree of cure (DoC) corresponding to Point V. $\gamma$ is a parameter within the limits -1 and 1. The value of $\gamma$ physically represents how rapidly the modulus develops initially until it reaches upper bound $\alpha_2$. Finally, $\alpha_{mod}$ is calculated as $\alpha_{mod} = \frac{\alpha - \alpha_1}{\alpha_2 - \alpha_1}$ such that it takes a value $0 \leq \alpha_{mod} \leq 1$.

## 2.4 Cure shrinkage model

An abrupt rise in the resin viscosity is generally considered representative of the crosslinking reactions in the resin leading to cure shrinkage. In the present study the cure shrinkage is assumed to occur in the range between point P, the minimum of viscosity and Point V, the vitrification point beyond which chemical reactions are assumed to cease (see Fig 2).

The resin volumetric shrinkage as a function of DoC is given by (Bogetti and Gillespie 1992)

$$V_r^s = \begin{cases} 0.0 & \alpha \leq \alpha_{c1} \\ A\alpha_s + (V_r^{s\infty} - A)\alpha_s^2 & \alpha_{c1} \leq \alpha \leq \alpha_{c2} \\ V_r^{s\infty}, & \alpha \geq \alpha_{c2} \end{cases} \quad (5)$$

Again, $\alpha_{c1}$ and $\alpha_{c2}$ are the DoC parameters for cure shrinkage model. Referring to Fig 2, $\alpha_{c1}$ is the DoC corresponding to Point P while, $\alpha_{c2}$ is to the DoC corresponding to Point V. Where $V_r^{s\infty}$ is the maximum volumetric shrinkage corresponding to 3% stain (Bogetti and Gillespie 1992), A is a constant and $\alpha_s$ is given as $\alpha_s = \frac{\alpha - \alpha_{c1}}{\alpha_{c2} - \alpha_{c1}}$. Resin shrinkage strains calculated from $V_r^s$ as:

$$\varepsilon_r^s = (1 + V_r^s)^{1/3} - 1 \quad (6)$$

The stress in the composite laminate calculated at the kth time increment is given by the following constitutive relationship:

$$\sigma_i^k = \sigma_i^{k-1} + \Delta\sigma_i^k = \sigma_i^{k-1} + \sum_{j=1}^{6} C_{ij}^k \Delta\varepsilon_j^{eff,k} \quad (i,j = 1-6) \quad (7)$$

Where $C_{ij}^k$ are the stiffness coefficients in the kth increment, and $\varepsilon_j^{eff,k}$ is the effective strain calculated by:

$$\Delta\varepsilon_i^{eff} = \Delta\varepsilon_i^{total} - \Delta\varepsilon_i^{th} - \Delta\varepsilon_i^{csh} = \Delta\varepsilon_i^{total} - CTE_i\Delta T - CCS_i\Delta\varepsilon_r^s \quad (8)$$

Where $\varepsilon_i^{total}$ is the total mechanical strain and the non-mechanical strains $\varepsilon_i^{th}$, $\varepsilon_i^{csh}$ are thermal and cure shrinkage strains respectively. $CTE_i$ is the coefficient of thermal expansion and $CCS_i$ is the coefficient of cure shrinkage. The stiffness coefficients $C_{ij}$, thermal coefficients $CTE_i$ and cure shrinkage coefficients $CCS_i$ are functions of constituent fiber and matrix properties and their effective composite properties are determined by self-consistent micromechanical homogenization equations (Scott R. White and Kim 1998).

## 2.5 Black-box optimization techniques

Genetic Algorithm NSGA-II is typically used to solve the cure optimization problem. Genetic Algorithms (GA), in general, overcome the limitations of the classical techniques (direct or gradient-based algorithms) such as dependence on the chosen initial solution and getting stuck to a suboptimal solution. Among the different genetic algorithms, NSGA-II was employed in this work for its key features including: (1) elitism, where top individuals are carried to the next generation, (2) crowding distance sorting to maintain diversity, and (3) non-dominated sorting, which ranks solution fronts based on dominance, prioritizing higher-ranked fronts for the next generation (Deb et al. 2002).

## 3. Cure Process Optimization with constrained Bayesian Optimization

Bayesian optimization (BO) (Frazier 2018) proficiently handles the optimization of black-box functions that are expensive or lack analytical expressions by iteratively developing a statistical model, typically a Gaussian process (GP) (Rasmussen and Williams 2005), capturing uncertainty to guide the optimization process. Bayesian optimization efficiently manages the trade-off between exploring new areas and exploiting known regions to optimize an objective function. Bayesian optimization process iterates over two steps: statistical model update and acquisition function optimization. Given by literature (Gardner et al. 2014) , statistical models should be customized by the optimization problem. A general framework of BO is reviewed in Appendix A. In our problem, we use Gaussian process models as surrogate models for objective function and constraint function and employ expected constrained improvement as the acquisition function.

Consider the problem that
$$\begin{aligned} min \quad & f(x) \\ s.t. \quad & g(x) \geq c \\ & x \in \mathcal{X} \end{aligned}$$
where input $x$ is $d$ dimensional, encompassing variables such as time and temperature. $c$ represents a constant, serving as the criteria of degree of cure in our cases. The functions $f(x)$ and $g(x)$ correspond to the objective function Deformation $u$ and the constraint function Degree of Cure (DoC), respectively, and both are black-box functions. To handle black-box functions in cure process, we fit GP model separately for each of these functions (Rasmussen and Williams 2005).

Consider the deterministic response $y(x)$ as a realization of a Gaussian stochastic process
$$Y(x) = \mu + Z(x) \tag{9}$$
where $\mu$ is the constant mean, and $Z(x)$ is a zero-mean, stationary, Gaussian stochastic process with variance $\sigma^2$, and correlation function $r(x, x')$. The correlation function is also known as a kernel function in the context in GP, and it defines how the influence of a single observation propagates throughout the input space, affecting predictions at other points.

We assume that the objective function $f(\cdot)$ and the constraint function $g(\cdot)$ are realizations of two independent Gaussian processes $Y^f(x)$ and $Y^g(x)$ defined similarly as in Eq. (9). Let $X_n = \{x_1, \cdots, x_n\}$ be the initial input points, and $Y_n^f = (y_1^f, \cdots, y_n^f)^\top$ and $Y_n^g = (y_1^g, \cdots, y_n^g)^\top$ be the corresponding outputs of the objective function $f(\cdot)$ and the constraint function $g(\cdot)$. Then for a new input point $x_{n+1}$, we have that
$$Y^f(x_{n+1})|Y_n^f \sim N(\hat{y}^f(x_{n+1}), s^{f,2}(x_{n+1})) \tag{10}$$

$$Y^g(x_{n+1})|Y_n^g \sim N(\hat{y}^g(x_{n+1}), s^{g,2}(x_{n+1})) \quad (11)$$

where

$$\hat{y}^f(x_{n+1}) = \hat{\mu}_f + \boldsymbol{r}_f^\top \boldsymbol{R}_f^{-1}(Y_n^f - \mathbf{1}\hat{\mu}_f), \text{ and } s^{f,2}(x_{n+1}) = \hat{\sigma}_f^2\left[1 - \boldsymbol{r}_f^\top \boldsymbol{R}_f^{-1}\boldsymbol{r}_f + \frac{(1-\mathbf{1}^\top R_f^{-1} r_f)^2}{\mathbf{1}^\top R_f^{-1}\mathbf{1}}\right];$$

$$\hat{y}^g(x_{n+1}) = \hat{\mu}_g + \boldsymbol{r}_g^\top \boldsymbol{R}_g^{-1}(Y_n^g - \mathbf{1}\hat{\mu}_g), \text{ and } s^{g,2}(x_{n+1}) = \hat{\sigma}_g^2\left[1 - \boldsymbol{r}_g^\top \boldsymbol{R}_g^{-1}\boldsymbol{r}_g + \frac{(1-\mathbf{1}^\top R_g^{-1} r_g)^2}{\mathbf{1}^\top R_g^{-1}\mathbf{1}}\right];$$

and $\hat{\mu}_f = \frac{\mathbf{1}^\top R_f^{-1} Y_n^f}{\mathbf{1}^\top R_f^{-1}\mathbf{1}}$, $\hat{\sigma}_f^2 = \frac{(Y_n^f - \mathbf{1}\hat{\mu}_f)^\top R_f^{-1}(Y_n^f - \mathbf{1}\hat{\mu}_f)}{n}$, and $\hat{\mu}_g = \frac{\mathbf{1}^\top R_g^{-1} Y_n^g}{\mathbf{1}^\top R_g^{-1}\mathbf{1}}$, $\hat{\sigma}_g^2 = \frac{(Y_n^g - \mathbf{1}\hat{\mu}_g)^\top R_g^{-1}(Y_n^g - \mathbf{1}\hat{\mu}_g)}{n}$.

$\boldsymbol{r}_f$ and $\boldsymbol{r}_g$ are vectors of correlations $(r_f(x, x_1), \cdots, r_f(x, x_n))^\top$ and $(r_g(x, x_1), \cdots, r_g(x, x_n))^\top$, $\boldsymbol{R}_f$ and $\boldsymbol{R}_g$ are the correlation matrices of size $n \times n$ with the $ij$th entry $r_f(x_i, x_j)$ and $r_g(x_i, x_j)$, respectively. More detailed forms are listed in Appendix B.

Expected Improvement (*EI*) (Jones, Schonlau, and Welch 1998) is one such acquisition function that measures the expected amount of improvement in the objective function value over the current best-known value at a given point. For our problem which is a minimization problem, *EI* at a point $x_{n+1}$ is defined as

$$EI(x_{n+1}) = \mathbb{E}(max\{0, Y_{\min} - Y^f(x_{n+1})\}|Y_n^f), \quad (12)$$

where $Y_{\min}$ is the current minimum value of the objective function, i.e., $Y_{\min} = \min(y_1^f, \cdots, y_n^f)$. And under GP, the closed form of *EI* can be expressed as

$$EI(x_{n+1}) = (Y_{\min} - \hat{y}^f(x_{n+1}))\Phi\left(\frac{Y_{\min} - \hat{y}^f(x_{n+1})}{s^{f,2}(x_{n+1})}\right) + s^{f,2}(x_{n+1})\phi\left(\frac{Y_{\min} - \hat{y}^f(x_{n+1})}{s^{f,2}(x_{n+1})}\right) \quad (13)$$

where $s^{f,2}(x_{n+1}) \neq 0$, $\phi(\cdot)$ and $\Phi(\cdot)$ are the standard normal density and distribution function.

In cure process cases containing a constraint in the optimization problem, a feasible solution should satisfy the constraint that $\hat{y}^g(x_{n+1}) \geq c$. Then a simple univariate Gaussian cumulative distribution function (Gardner et al. 2014), $PF(\cdot)$, is employed to weight for *EI* that

$$PF(x_{n+1}) \coloneqq Pr(\hat{y}^g(x_{n+1}) \geq c) = 1 - \Phi\left(\frac{c - \hat{y}^g(x_{n+1})}{s^{g,2}(x_{n+1})}\right) \quad (14)$$

Therefore, the Expected Constrained Improvement ($EI_C$) from (Gardner et al. 2014) will be introduced as acquisition function in our case that

$$EI_C(x_{n+1}) = EI(x_{n+1}) \cdot PF(x_{n+1}) \quad (15)$$

In our experiment, the next query point is selected from a candidate pool $\mathcal{D}$ in size $m \times d$ that $m$ samples are generated by Latin Hypercube Sampling (LHS) method, i.e., $\mathcal{D} = [d_1, \cdots, d_m]^\top$. Following Eq. (15), the next design point $x_{n+1}$ is derived such that

$$x_{n+1} = \arg \max_{i=1,\dots,m} EI_C(d_i) \quad (16)$$

The algorithm of Bayesian Optimization with Constraints is shown in Table 1.

Table 1. Algorithm: Bayesian Optimization with Constraints

---

**Algorithm 1** Bayesian Optimization with Constraints

1: **Preliminaries**: Objective function $f$, constraint functions $g$, acquisition function $EI_C$, maximum iterations $N$.
2: **Input**: Initial data set $X_n = \{x_1, \cdots, x_n\}$, $Y_n^f = (y_1^f, \cdots, y_n^f)^\top$, $Y_n^g = (y_1^g, \cdots, y_n^g)^\top$.
3: **Output**: The best point $x^*$ found.
4: Initialize data $X_n, Y_n^f, Y_n^g$, and the current optimal point $x^* = \arg\min_{\substack{x \in X_n \\ g(x) \geq c}} f(x)$
5: **for** $i = n$ to $N$ **do**
6:     Update the GP models for $f$ and $g$ based on data $X_i$.
7:     Generate a random candidate pool using the LHS method to determine the new design point $x_{i+1}$, as described in Equation (16).
8:     Evaluate the true objective function and constraint function values, $y_{i+1}^f = f(x_{i+1})$ and $y_{i+1}^g = g(x_{i+1})$.
9:     By including new point $x_{i+1}$, $y_{i+1}^f$, and $y_{i+1}^g$, update $X_i \to X_{i+1}$, $Y_i^f \to Y_{i+1}^f$, and $Y_i^g \to Y_{i+1}^g$.
10:     **if** no constraints are violated and $y_{i+1}^f < f(x^*)$ **then**
11:         Update the best point $x^* = x_{i+1}$
12:     **end if**
13: **end for**
14: **return** $x^*$

---

## 4. Implementation of Cure Process Optimization Approaches through Case Studies

The two optimization approaches described in Section 2 were implemented for each of two case studies that have different cure scenarios and geometric configurations. The first case study includes cure of a flat rectangular laminate with a cross-ply layup that leads to out-of-plane PID at the end of cure. The second case study consists of the cure of L-shaped laminate that leads to spring-in PID and twisting (rotational)PID at the end of cure. The cure problem for each case study including the dimensions and layup of the geometry, inputs and boundary conditions for the cure process model, optimization setup and cases are detailed in this section.

### 4.1 Case Study 1: Cure of Flat Rectangular Laminate

As described in Section 2, the cure process model consists of two separate model setups for the thermo-chemical step and the stress-deformation step that are sequentially coupled for cure analysis. Accordingly, the two model setups were created for the cure analysis of the flat rectangular laminate. First, the computational model was first validated with a study in literature (Shah et al. 2018). The model was implemented for a flat rectangular laminate of dimensions 152mm×25mm×1.2mm with $[0_4/90_4]$ layup and two-step cure cycle shown in Fig 2 with the degree of cure curves from the present work[33] exhibiting good agreement with the results given in Ref. (Shah et al. 2018). Further, the asymmetric layup considered in this study produced an out-of-plane curvature as obtained at the end of the stress-deformation analysis. The curvature obtained from the present work (42.42e-4) had good agreement (difference = 3.4%) with the results of Ref. (Shah et al. 2018) (43.92e-4).

Next, the validated cure process model was used for analysis of the flat rectangular thin laminate geometry with dimensions 152mm×25mm×0.3mm and layup [0/90]. In the Thermo-chemical model setup, the cure cycle was assigned to all the surfaces of the laminate by defining a temperature boundary condition (BC). The baseline cure cycle was defined with a ramp up rate of 2.6 °C/min, a dwell at 180°C for 112 min and a cool down rate of 4.846 °C/min. The stress/deformation predictions from the cure process model defined with the baseline cure cycle were considered the benchmark for comparison of results. The laminate was defined with an initial temperature BC of 20°C. Note that convection heat transfer with the surrounding air was not considered in the study. A mesh of 20-node solid elements (C3D20) was defined, and a transient analysis was performed. The results obtained at the end of the thermo-chemical analysis namely temperature and degree of cure over the entire laminate were provided as inputs to the stress-deformation analysis. The structural BCs in the stress-deformation step were defined as shown in Fig 3.

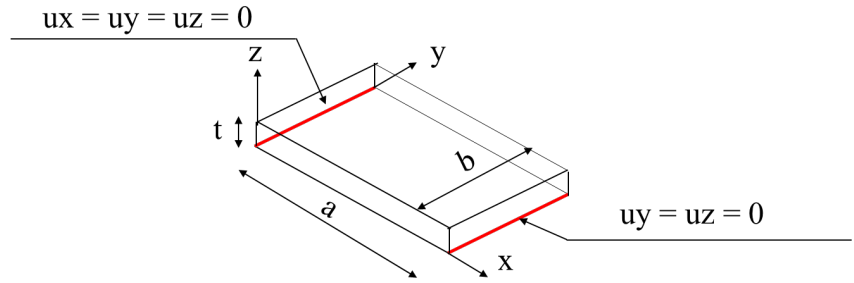

Figure 3. Structural boundary conditions defined in the stress-deformation step.

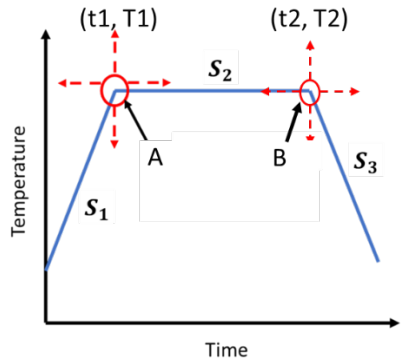

Figure 4. Cure cycle parameters: coordinates of point A (t1, T1) and B (t2, T2) used for optimization study.

The design input variables of the cure optimization problem were the coordinates (*t1, T1*) of the point A on the baseline cure cycle shown in Fig 4. The location of this point in the *t-T* plane controlled the slopes of segments $S_1$, and $S_2$ as well as the interaction between thermal and cure shrinkage effects. The constraints imposed on the optimization process ensured that (1) cured composite structure is acceptably cured (DoC ≥ 0.990) and (2) modified cure cycle designs always result in interaction of the cure shrinkage and thermal effects to reduce PID ($slope(S_1) > slope(S_2)$). The two cure optimization cases R1 and R2 were considered as shown in Table 2. The selection of the cases was based on two considerations: (1) as input variable *t* is the time for heating the composite laminate in the cure cycle, two values of *t1*=1 min for case R1 and *t1*=10min for case R2 were selected for evaluation considering the manufacturing efficiency and feasibility. An upper bound of 110 min was defined for *t1* whereas *T1* was defined such that 125<= *T1* <=180, since the design samples outside these bounds had shown to be infeasible in trial runs; (2) the final value of DoC must be higher than 0.990 which is considered full cure of the resin for sufficient modulus development and final properties of the composite structure. Hence DoC=0.995 for cases R1, R2 were considered for evaluation of the optimized cure cycle. In the black-box GA optimization, the initial population/ designs were defined to be 100. The optimization algorithm was run for 10 generations to perform a total of 1000 evaluations.

Table 2. Optimization cases performed for the flat rectangular laminate

| Cases | DoC | Range of *t1* (min) | Range of *T1* (°C) |
|---|---|---|---|
| R1 | ≥ 0.995 | 1-110 | 125-180 |
| R2 | ≥ 0.995 | 10-110 | 125-180 |

In Bayesian optimization implementation, we select 10 initial samples from the flat rectangular laminate numerical experiments. Given that in real-world scenarios of flat laminates, slopes of $S_1$ and $S_2$ are non-negative with $slope(S_1) > slope(S_2)$, we omit the slope constraints in the Bayesian optimization approach. Surrogate models of *u* and DoC are constructed respectively based on these initial samples. Subsequently, a candidate pool in size of 10,000 is established, from which predictions of u and DoC are made using the constructed surrogate models. The next experimental point ($t_{new}$, $T_{new}$) is chosen which is the candidate possessing the highest expected constrained improvement value among all candidates, as delineated by Equation (18). The *u* and DoC values corresponding to this new point ($t_{new}$, $T_{new}$) are derived from the cure of flat rectangular laminate experiment, and this new sample is added to the current set to update the surrogate models. With a budget of 40 steps for the flat rectangular laminate experiment—10 for initial sampling and 30 for deriving results through constrained Bayesian optimization. The minimum u over the last 30 steps, associated DoC meeting the DoC constraint, is considered as the optimal value.

## 4.2 Case Study 2: Cure of L-shaped laminate

Similar to case study 1, a cure process model consisting of two model setups for thermo-chemical and stress-deformation analyses respectively was developed for L-shaped laminate. The dimensions of the L-shaped laminate are shown in Fig 5(a). The cure cycle was defined as thermal boundary condition in the thermos-chemical step and the mechanical boundary conditions were defined in the stress-deformation step as shown in Fig 5(b). A uniform pressure of 7MPa is applied to the upper surface to mimic the forming process of the laminate during the entire cure period in the stress-deformation step and was removed at the demolding step. Since the L-shaped laminate is symmetric about the plane in the curved section of the laminate shown in the figure, only one-half of the structure was modeled.

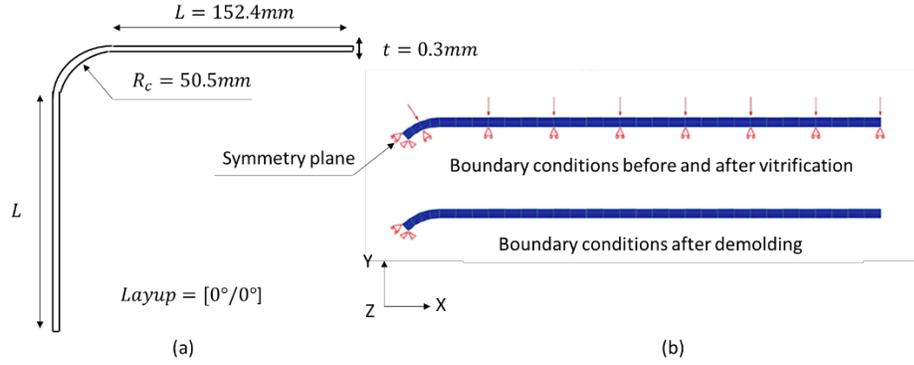

Figure 5. (a) Geometry and (b) mechanical boundary conditions of L-shaped laminate (Zhang et al. 2019)

The design input variables of the cure optimization problem were the coordinates of Point A (*t1, T1*) and point B (*t2, T2*) on the baseline cure cycle shown in Fig 4. The location of these points in the *t-T* plane controlled the slopes of segments $S_1$, and $S_2$ as well as the interaction between thermal and cure shrinkage effects. The upper and lower bounds of time variables were defined as 10<= *t1* <=110, and 120<= *t2* <=200, while for the temperature variables, the bounds were defined as 125<= *T1* <=180 and 150<= *T2* <=180, since the design samples outside these bounds had shown to be infeasible in trial runs.

Two optimization cases namely Q1, and Q2 were considered as shown in Table 3. The selection of the cases and the imposed constraints were based on the following considerations: (1) expansion of the design space by using 4 variables *t1, T1, t2, T2* for Q1 and Q2 cases compared to the two variables *t1, T1* for the R1 and R2 cases for the Flat rectangular laminate, (2) change of layup in composite laminates significantly changes the stress/deformation response in the demolding step after cure, thus two layup configurations [0°/0°] for case Q1 and [45°/-45°] for case Q2 were studied, (3) considering cure at a lower temperature may appreciably change the stress/deformation response, a case Q1 with constraint DoC > 0.960 and other case Q2 with full cure were studied, and finally, (4) the constraint ($slope(S_1) > slope(S_2)$) ensures interaction of the cure shrinkage and thermal effects to reduce PID hence is defined for both cases Q1 and Q2. However, while $slope(S_1)>0$ is always true because $S_1$ is the heating phase, $slope(S_2)$ may be positive or negative that changes stress/deformation response. Thus, the constraint $slope(S_2) > 0$ was defined for case Q2 while Q1 was allowed to use $\pm slope(S_2)$. In the black-box GA optimization, the initial population/ designs were defined to be 100. The optimization algorithm was run for 10 generations to perform a total of 1000 evaluations.

Table 3. Optimization cases performed for L-shaped Laminate

| Case | Layup | DoC constraint | $slope(S_2)$ constraint |
|---|---|---|---|
| Q1 | [0°/0°] | ≥ 0.960 | - |
| Q2 | [45°/-45°] | ≥0.990 | > 0.0 |

In the study on the cure of L-shaped laminate, akin to the Bayesian optimization approach employed in the flat rectangular laminate experiment, we commence by selecting 15 initial samples. Subsequently, surrogate models for u and DoC are established separately based on these samples, represented as $\hat{u}$ (*T1,T2,t1,t2*) and $\widehat{DoC}$ (*T1,T2,t1,t2*). Upon generating a candidate pool comprising 10,000 candidates, we proceed to sieve through the candidates, retaining those that meet the slope constraints, and derive predictions for u and doc for the filtered candidates. Same as the methodology applying Eq. (18) in the flat rectangular laminate experiment, we determine the subsequent sampling point. The newly acquired sample is integrated into the existing sample set, thereby preparing the refreshment of the surrogate models. This iterative process is executed a total of 35 times, culminating in the selection of the first minimum value as the optimal, culminating in the identification of the minimum u value as the optimal, ensuring that the corresponding DoC also meets the requisite DoC constraint.

## 5. Results and Discussion

### 5.1 Numerical validation of Bayesian optimization

We first perform numerical verification of the constrained Bayesian optimization approach via a simplified analytical PID model. We use the second-order polynomial regression to construct the *u* and DOC functions based on existing data collected from case study 1. The fitted polynomial model allows us to replicate the data generate process in Fig 1 efficiently and validate the performance of cBO through a large number of micro-replications.

In numerical verification experiment, time (*t*) ranges between 1 and 108 minutes, the temperature (*T*) varies within a range from 120 to 177 degrees Celsius, and degree of cure (DoC) is greater than 0.995. To simplify subsequent processes, we normalize the input variables *t* and *T* to the uniform interval [0,1]. Therefore, the optimizing problem is

$$\begin{aligned}
\min \quad & u(t,T) = -0.1272t^2 - 0.1698tT + 0.2914t + 0.2329T^2 - 0.0841T + 1.8646 & (17)\\
s.t. \quad & DoC(t,T) = -0.0458t^2 + 0.0801tT - 0.0265t - 0.0376T^2 + 0.0329T + 0.9902 \geq 0.995 \\
& 0 \leq t \leq 1 & (18)\\
& 0 \leq T \leq 1
\end{aligned}$$

The objective function Eq. (17) exhibits neither convex nor concave characteristics, whereas the constraint function Eq. (18) is identified as convex. Thus, a closed-form solution is not available for this problem system. To address this challenge, we implement cBO strategy, seeking a feasible solution to this problem.

Fig 6 displays the initial 50 steps of the cBO process in numerical verification experiment. The depicted trend reveals a swift decline in the median, the 5th percentile, and the 95th percentile within the first 15 steps. Subsequently, all three measures continue their convergence towards the minimum value. Table 4 shows that all those three measures stabilize after 50 steps with mean 1.8570, the 5th percentile 1.8570 and the 95th percentile 1.8571.

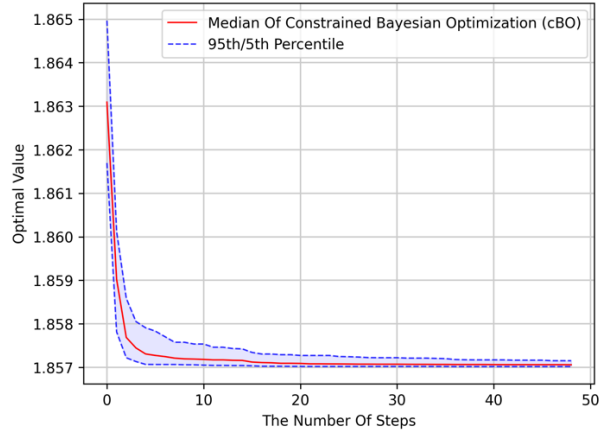

Figure 6. cBO performance in the numerical validation example.

Table 4. Constrained Bayesian Optimization in numerical validation example: results of 100 Steps

| The Number of Steps | Mean | Medium | 5th Percentile | 95th Percentile |
|---|---|---|---|---|
| 10 | 1.8572 | 1.8572 | 1.8571 | 1.8575 |
| 20 | 1.8571 | 1.8571 | 1.8570 | 1.8573 |
| 30 | 1.8571 | 1.8571 | 1.8570 | 1.8572 |
| 50 | 1.8571 | 1.8571 | 1.8570 | 1.8572 |
| 75 | 1.8571 | 1.8570 | 1.8570 | 1.8571 |
| 100 | 1.8570 | 1.8570 | 1.8570 | 1.8571 |

Within the GA, the process has generated 5 generations, resulting in an accumulation of 1490 function evaluations. The median decreases from 1.8627 to 1.8574, 5th percentile declines from 1.8572 and 1.8570, and 95th percentile descends from 1.8808 to 1.8646. However, cBO approach reaches its minimum result 1.8570 in approximately 75 steps with a lower minimum value than GA 1.8574.

Table 5 illustrates the performance of cBO over the initial 100 steps and the performance of GA between the 150th and 650th function evaluations. It is evident that cBO achieves convergence towards a better optimal value within the first 100 steps than the optimal value attained by GA between the 150th and 650th steps. Additionally, the range between the 5th percentile and the 95th percentile of cBO is notably narrower compared to that of GA.

Table 5. Comparative results in Example 1 from Genetic Algorithm and constrained Bayesian Optimization

| Approach | The Number of Steps | 5th Percentile | Optimal Value | 95th Percentile |
|---|---|---|---|---|
| GA | 150 | 1.8572 | 1.8638 | 1.8819 |
|  | 344 | 1.8570 | 1.8597 | 1.8740 |
|  | 450 | 1.8570 | 1.8590 | 1.8761 |
| cBO | 30 | 1.8570 | 1.8571 | 1.8572 |

**5.2 Comparative results of GA and Bayesian cure process optimization**

### 5.2.1 Case study 1

Fig 7 shows comparative plots of convergence for the R1 and R2 optimization cases performed for the Flat rectangular laminate. The plot shows convergence of deformation $u$ over the model evaluation steps. The orange curve shows convergence of GA while the blue curve shows convergence of the constrained Bayesian optimization (cBO). It is evident from the plots that while the GA continues to have oscillations even after 1000 iterations, the cBO converges to a stable $u$ value in less than 40 steps.

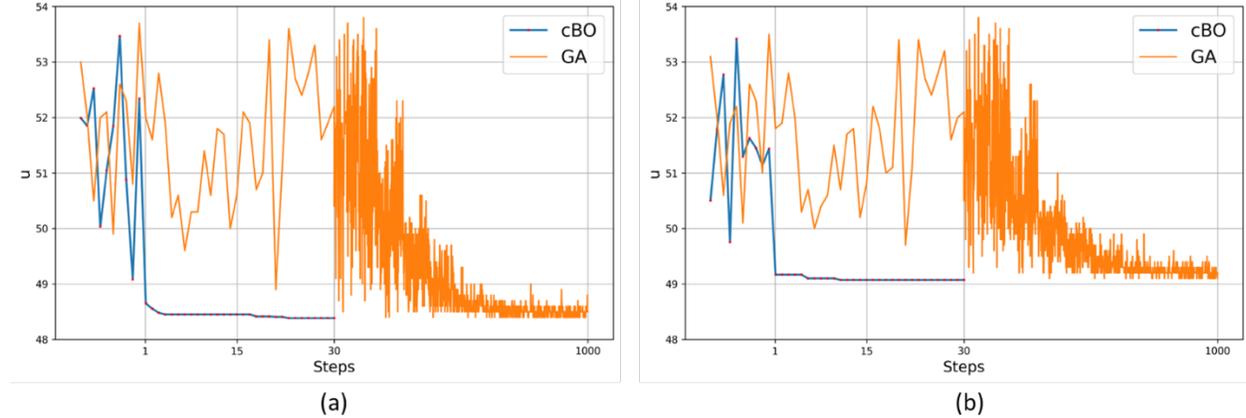

Figure 7. Comparative plot of convergence for (a) R1 and (b) R2 optimization cases

The comparison of optimum cure cycle parameters obtained from GA and cBO for the R1 and R2 cases are provided in Table 6. Further, the results of optimization namely the deformation $u$ and the final DoC are provided in Table 7. It is observed that the results from the GA and cBO approaches agree very well suggesting that the cBO is valuable in solving the considered cure problem. The comparison of cure cycle parameters ($t1, T1$) shows only one notable discrepancy in $t1$ parameter for R1 case. This suggests that the results are not sensitive to the $t1$ parameter. This can be explained by the fact that the early heating phase does not contribute to the development of residual stresses/deformation, thus a difference of <30% in $t1$ does not affect the result values. Table 7 also provides computational efficiency of cBO compared to the GA which is approximately 96% considering the GA takes more than 1000 iterations while cBO converges within 40 iterations.

Table 6. Comparison of Optimum cure cycle parameters for R1 and R2 cases

| Case | $t1(min)$ | | % error | $T1$ (°C) | | % error |
|---|---|---|---|---|---|---|
| | GA | cBO | | GA | cBO | |
| R1 | 1.001 | 1.281 | 27.972 | 134.36 | 134.58 | 0.164 |
| R2 | 10.23 | 10.19 | 0.391 | 135.76 | 135.37 | 0.287 |

Table 7. Comparison of deformation $u$ and DoC for R1 and R21 cases and estimate of computational efficiency of cBO

| Case | $u$ (mm) | | DoC | | Convergence | Computational |

|    |       |       | %     |       |       | %     | steps  |      | Efficiency of |
|----|-------|-------|-------|-------|-------|-------|--------|------|---------------|
|    | **GA** | **cBO** | error | **GA** | **cBO** | error | **GA** | **cBO** | cBO (%) |
| R1 | 48.36 | 48.38 | 0.041 | 0.996 | 0.996 | 0 | > 1000 | < 40 | 96 |
| R2 | 49.06 | 49.07 | 0.020 | 0.995 | 0.995 | 0 | > 1000 | < 40 | 96 |

Fig 8 shows contour plots of the flat rectangular laminate subjected to the baseline and optimized cure cycles. The cross ply [0/90] layup used for the cure of the flat laminate produces an out-of-plane (along z-axis) deformation which is representative of the residual stresses developed in the laminate during cure.

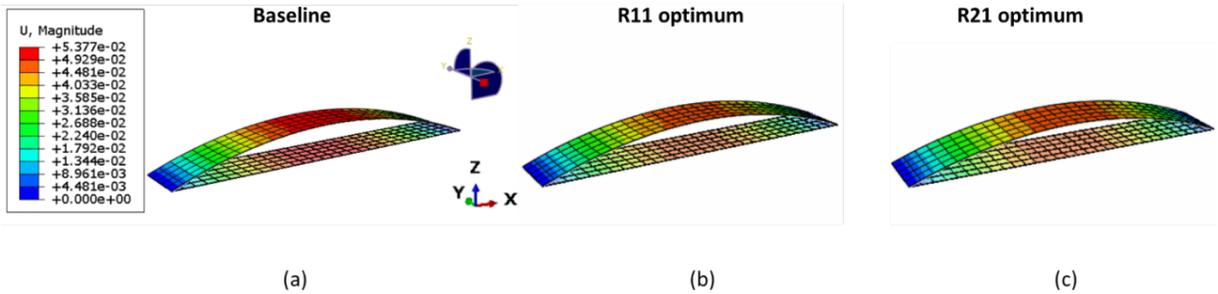

Figure 8. Contour plots of deformation u for Flat rectangular laminate subjected to (a) Baseline, (b) R1 optimum and (c) R2 optimum cure cycle

The baseline and optimum cure cycles for the R1 and R2 cases are shown in Fig 9(a). The corresponding deformation curves are shown in Fig 9(b). It is observed that the optimum cure cycles maximize the period of the second heating phase such that the interaction of the Thermal and Cure shrinkage effects is maximized. For baseline case, the cure shrinkage effects are dominate from Gelation to Vitrification (V) whereas for the optimized cases these effects are dominant for a prolonged time from G to V' due to slower heating rate. Thus, the cure shrinkage effects that cause an increase in deformation from G to V in the baseline case are counteracted by the thermal effects in the optimized cases. This is visible in the deformation curves of the optimized cases that have a lower slope as compared to the baseline cases. The deformation magnitude is further reduced in the optimized cases by the thermal expansion effects dominant up to the cooling phase. Thus, the overall reduction in deformation for the optimized cure cycle cases R1 and R2 compared to the baseline is a considerable 9-10%.

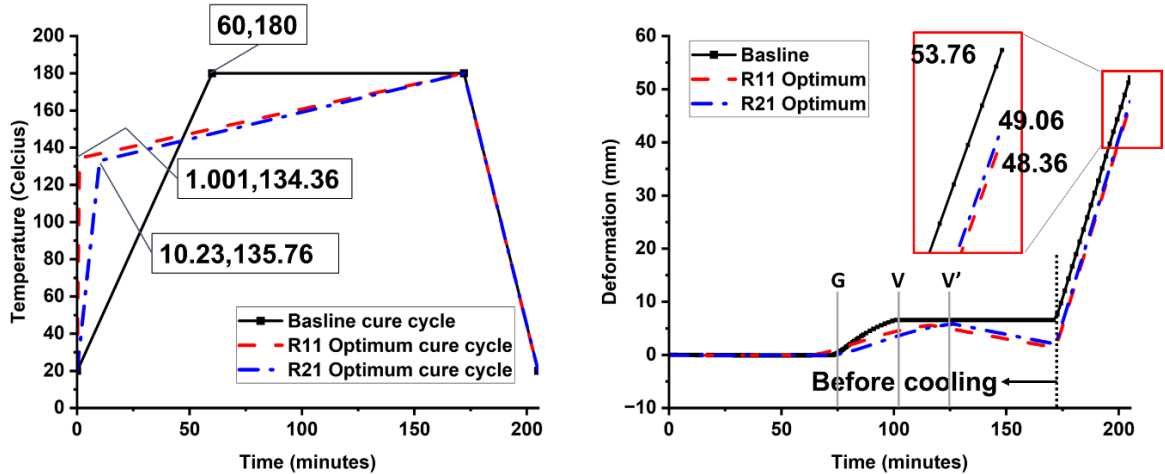

Figure 9. (left) Baseline and Optimized cure cycles, (right) deformation curves for the baseline and optimized cure cycles

**5.2.2 Case study 2**

Fig 10(a) shows the outcomes of applying the GA and cBO approaches to case Q1 of the cure process of the L-shaped laminate. GA was executed for 1003 iterations to reach convergence, while cBO is limited to 50 steps in total. Within the cBO framework, the first 15 points were derived from preliminary samples of L-shaped laminate numerical experiments, and the subsequent 35 points constituted the learning phase of Bayesian optimization. GA showed the ability to converge to the optimal value only after performing a considerable number of iterations, while cBO showed extremely fast convergence, achieving the optimal value in 35 steps of learning.

Fig 10(b) illustrates a comparative analysis of outcomes of case Q2 using the GA and cBO approaches. The GA was executed for 1113 iterations, whereas the cBO was constrained to 50 steps. Within these, the initial 15 data points preceding Step 1 on the cBO trajectory were obtained from the L-shaped laminate experiment, succeeded by 35 steps of the cBO learning process. Notably, in case Q2 of the L-shaped laminate, the outcome $u$ exhibits pronounced oscillations, accounting for acute vertices appearing in the GA curve prior to convergence. The GA method, albeit demonstrating evident convergence within 1113 steps, continues to manifest oscillatory behavior at the end of GA execution, suggesting the need for additional iterations to attain convergence to an optimal state. Conversely, the Bayesian Optimization method exhibits a rapid rate of convergence, achieving an optimal value within the span of 35 steps.

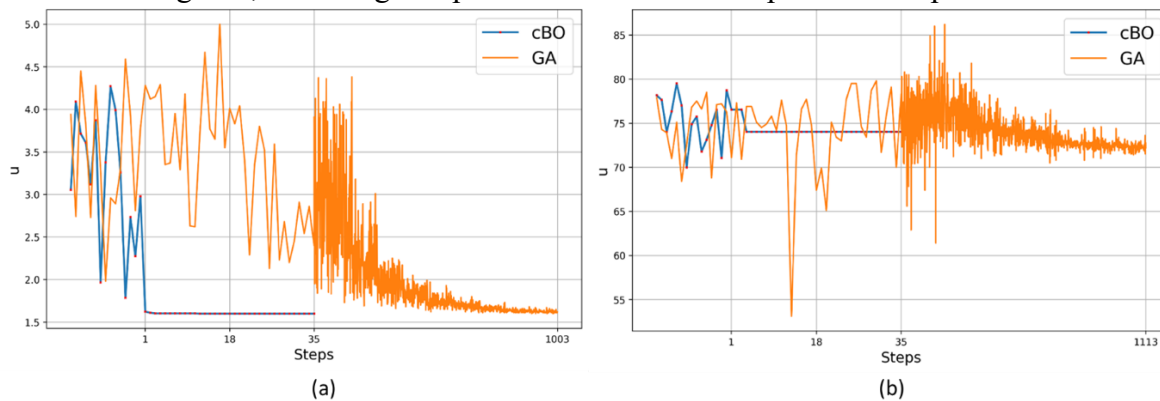

Figure 10. Comparative plot of convergence for (a) Q1 and (b) Q2 optimization cases

The cure cycle parameters obtained from GA and cBO for the L-shaped laminate cases are provided in Table 8 and the results $u$ and DoC are provided in Table 9. The GA and cBO results for Q1 have good agreement with and a 96% computational efficiency of cBO. The Q2 case shows higher prediction discrepancy in parameter $t1$ with a maximum error of 26.3%. However, the corresponding error in $u$ is < 4% indicating the cure cycle parameter $t1$ is not sensitive to the deformation results, which is also observed in the flat laminate case. The computational efficiency was calculated based on the number of steps/iterations of convergence for GA and cBO. The computational efficiency of cBO for Q1 was calculated to be 96%. For case Q2, the GA iterations to convergence is more than 2000 which is significantly higher than the previous cases. The lower convergence rate in GA can be attributed to increased complexity of the cure problem. The cBO approach for this case converged within 50 steps. Thus, the computational efficiency (97.5%) was higher as compared to the other cases suggesting that the cBO approach is valuable for more complex cure problem.

Table 8. Comparison of Optimum cure cycle parameters for Q1 and Q2 cases

| Case | $t1$ (min) | | % error | $T1$ (°C) | | % error | $t2$ (min) | | % error | $T2$ (°C) | | % error |
|---|---|---|---|---|---|---|---|---|---|---|---|---|
| | GA | cBO | | GA | cBO | | GA | cBO | | GA | cBO | |
| Q1 | 10.11 | 10.01 | 0.989 | 179.45 | 179.56 | 0.089 | 196.56 | 198.75 | 1.114 | 151.66 | 157.80 | 4.049 |
| Q2 | 10.15 | 13.47 | 26.261 | 142.17 | 125.97 | 11.395 | 197.96 | 185.47 | 6.309 | 163.07 | 178.97 | 9.750 |

Table 9. Comparison of deformation u and DoC for Q1 and Q2 cases and estimate of computational efficiency of cBO

| Case | $u$ (mm) | | % error | DoC | | % error | Convergence steps | | Computational Efficiency of cBO (%) |
|---|---|---|---|---|---|---|---|---|---|
| | GA | cBO | | GA | cBO | | GA | cBO | |
| Q1 | 1.598 | 1.6023 | 0.269 | 0.999 | 0.999 | 0 | > 1000 | < 40 | > 96 |
| Q2 | 71.58 | 74.02 | 3.409 | 0.990 | 0.991 | 0.101 | > 2000 | < 50 | > 97.5 |

Fig 11(a) and (b) show the deformation contour plots of the L-shaped laminate with [0/0] layup subjected to Baseline and Q1 optimum cure cycle respectively. Since the laminate is symmetric about the curved plane, results are displayed for only one half of the laminate. The deformation of this laminate is only in the thickness direction causing a spring-in of the L-structure. Similarly, Fig 11(c) and (d) show the deformation contours for the L-shaped laminate with [45/-45] layup subjected to baseline and Q2 optimum cure cycles respectively. The deformation for this L-shaped laminate consists of twisting modes which is observed in the contour plots (c) and (d).

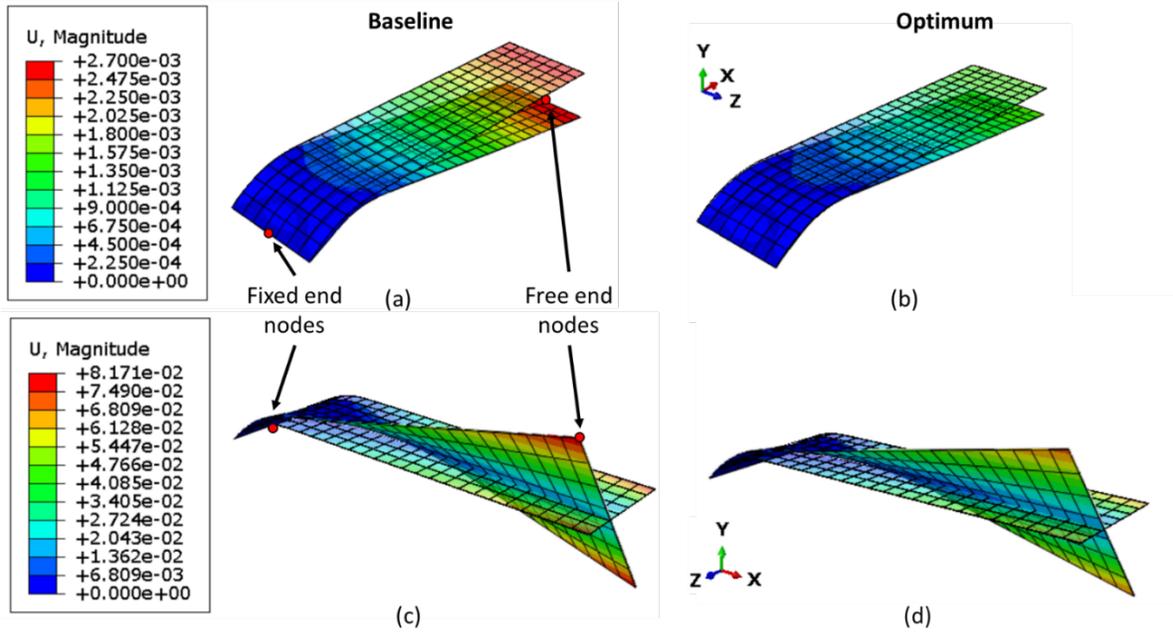

Figure 11. Deformation Contour plots for L-shaped laminate with [0/0] layup subjected to (a) baseline cure cycle, (b) Q1 optimum cure cycle and with [45/-45] layup subjected to (c) Baseline cure cycle and (d) Q2 optimum cure cycle

The baseline and optimum cures cycles for the Q1 and Q2 L-shaped laminate cases are shown in Fig 12(a) and (b). The S2 segment of both the optimum cure cycles are longer than the baseline cycles. This suggests that expanding the design space by using four input variables instead of just used two for the flat laminate cases does change the decision of the optimum cure cycle. While both the optimum cure cycles cause interaction between the thermal and cure shrinkage effects, the Q1 optimum cure cycle stands out as unique because this is the only case where a negative slope of S2 produces a reduction in the cure induced deformation.

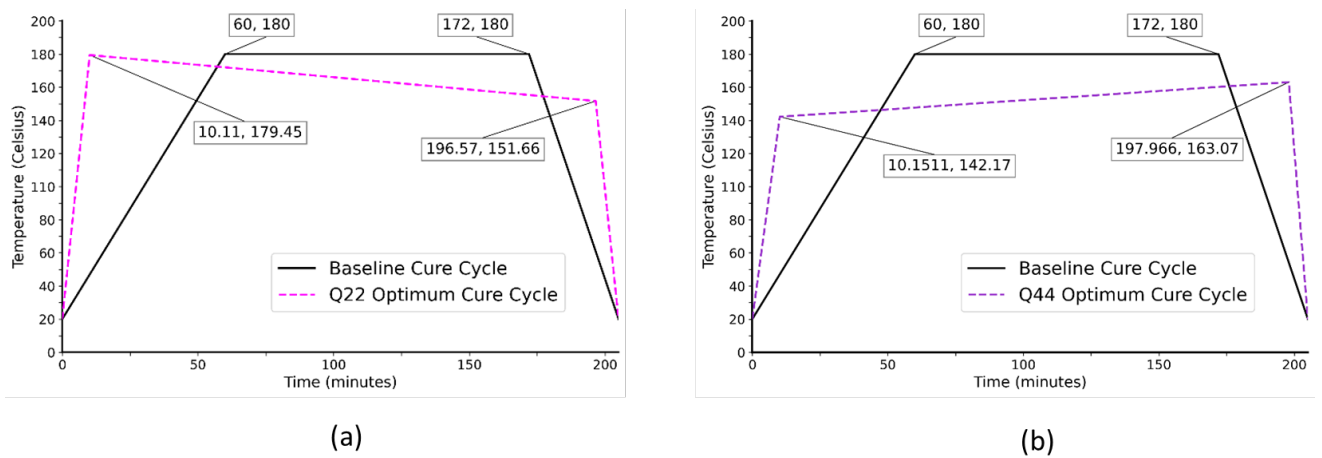

Figure 12. (a) Baseline and Q1 optimum cure cycle for L-shaped laminate with [0/0] layup, (b) Baseline and Q2 optimum cure cycle for L-shaped laminate with [45/-45] layup

Finally, Fig 13 shows the deformation plots for the Q1 and Q2 optimum cure cases along with their baseline cases. The deformation of the L-shaped laminate is constrained for the entire cure period through a uniform laminate pressure from the top and the tool on the bottom. As a result, the cure induced deformation is only evident in the demolding step when part of the residual stresses is released to allow the laminate to get to an equilibrium position through deformation.

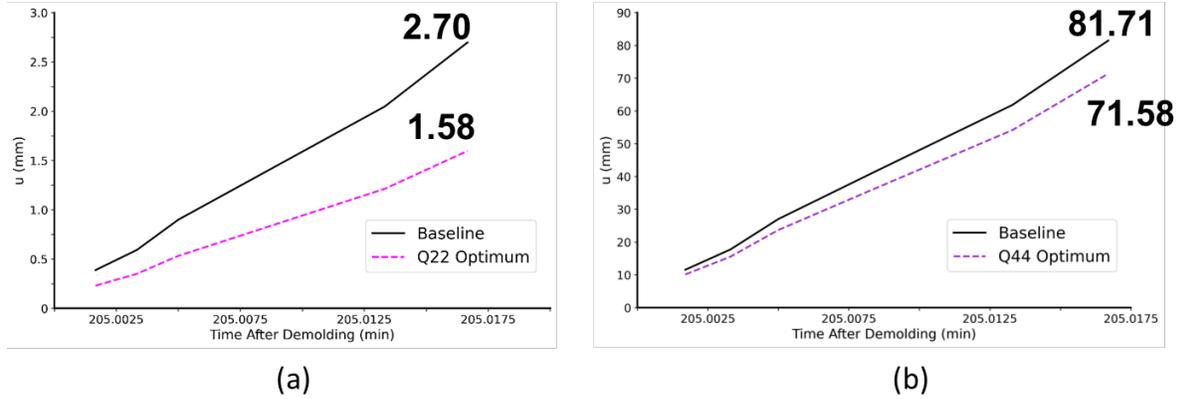

Figure 13. Laminate deformation on demolding for (a) baseline and optimum for case Q1, (b) baseline and optimum for case Q2.

The spring-in for the baseline cure cycle is 2.700mm while it reduces to 1.598mm with the Q1 optimum cure cycle. Thus, a significant (41.4%) reduction is achieved with the optimum cure cycle. While Q2 optimum cure cycle produces a cure deformation of 71.58mm as compared to the baseline of 81.71mm which amounts to a reduction of 12.4%.

## 6. Conclusion

In the present study, the cure optimization problem of laminated composites was solved through a statistical approach. The approach consisted of using constrained Bayesian Optimization (cBO) along with a Gaussian Process model as a surrogate to rapidly search for the optimal solution. The approach was implemented to two case studies including the cure of a simpler flat rectangular laminate and a more complex L-shaped laminate. The cure optimization problem with the objective to minimize cure induced distortion was defined for both case studies. The comparison of results from GA and cBO including deformation and final degree of cure showed good agreement (error < 4%). The computational efficiency of cBO for all optimization cases was found to be > 96% in comparison with the GA approach. It was concluded that the cBO approach was effective and much more efficient for solving the cure optimization problem. Further, the solution accuracy of cBO algorithm for multiple cure scenarios and geometric configurations established that the cBO approach can be reliably implemented for problems with larger size and geometric complexity.

**Acknowledgement**
Research supported as part of the AIM for Composites, an Energy Frontier Research Center funded by the U.S. Department of Energy (DOE), Office of Science, Basic Energy Sciences (BES), under Award #DE-SC0023389 (constrained Bayesian Optimization) and by the National Science Foundation (NSF) under Award # 2244342 (cure modeling).

## Appendix A:
## Bayesian Optimization Overview

Given a limited computational budget, Bayesian optimization (Frazier 2018) offers the advantage of efficiently optimizing black-box functions. The flowchart shown in Figure 3 presents Bayesian Optimization framework. Before starting the optimization, some initial data points are collected by sampling the function we aim to optimize across various input values. This initial data provides a starting point for the Bayesian optimization process. Using the initial data, a statistical model is constructed to represent our current beliefs about the black-box function. Then by maximizing the acquisition function, the next best location to sample the function is determined. After sampling the function at this new location, the data is updated, and the statistical model is refined. This process is repeated until a stopping criterion is met. The ultimate goal of Bayesian optimization is to find the input value that minimizes the expected value of the function given the most updated data.

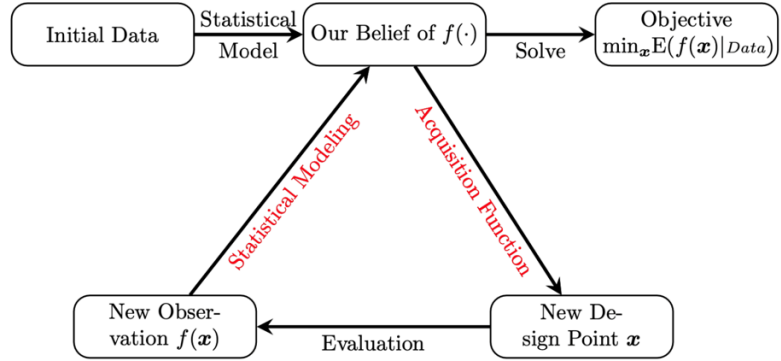

Figure 1. Bayesian Optimization Flowchart

## Appendix B
## Kernel Function

In our cases, we select the Matérn kernel with parameter 5/2 and a separate length scale per predictor ('ardmatern52' in Matlab *fitrgp* function) for both correlation functions $r_f(x_I, x_j)$ and $r_g(x_I, x_j)$ that

$$r_f(x_i, x_j) = \tau_f^2 \left(1 + \sqrt{5}r_d + \frac{5r_d^2}{3}\right) exp(-\sqrt{5}r_d),$$

$$r_g(x_i, x_j) = \tau_g^2 \left(1 + \sqrt{5}r_d + \frac{5r_d^2}{3}\right) exp(-\sqrt{5}r_d),$$

with

$$r_d = \sqrt{\sum_{h=1}^{d} \left(\frac{x_i^h - x_j^h}{l_h}\right)^2},$$

$$\tau_f^2 = \frac{sd(Y_n^f)}{\sqrt{2}},$$

$$\tau_g^2 = \frac{sd(Y_n^g)}{\sqrt{2}},$$

and the default initial value of the length scale parameters $l_h$'s are the standard deviations of the predictors that $l_h = sd(X_n^h)$, and $sd(\cdot)$ is standard deviation function.